\documentclass[10pt]{article}
\usepackage[english]{babel}
\usepackage{amsmath,amsthm}
\usepackage{amsfonts}
\usepackage{graphicx}
\usepackage{amssymb}
\usepackage[a4paper]{geometry}

\numberwithin{equation}{section}
\input{tcilatex}
\begin{document}

\title{The number $e^{\frac{1}{2}}$ is the ratio between the time of maximum
value and the time of maximum growth rate for restricted growth phenomena?}
\author{Zi-Niu WU\thanks{%
Corresponding Author, ziniuwu@tsinghua.edu.cn} \\
{\small Department of Engineering Mechanics, Tsinghua University, China}}
\date{October 1, 2013}
\maketitle

\begin{abstract}
For many natural process of growth, with the growth rate independent of size
due to Gibrat law and with the growth process following a log-normal
distribution, if the growth process is restricted so that the prodcution of
Shannon entropy is maximized, the ratio between the time (D) for maximum
value and the time (L) for maximum growth rate (inflexion point) is then
equal to the square root of the base of the natural logarithm ($e^{1/2}$).
On the logarithm scale this ratio becomes one half ($\frac{1}{2}$). It
remains an open question, due to lack of complete data for various cases
with restricted growth, whether this $e^{1/2}$ ratio can be stated as $%
e^{1/2}$-Law. \ Two established examples already published, one for an
epidemic spreading \ and one for droplet production, support however this
ratio. Another rough example appears to be the height of humain body. For
boys the maximum height occurs near 23 years old while the maximum growth
rate is at the age near 14, and there ratio is close to $e^{-1/2}$.

{\small \ }
\end{abstract}


\section{Natural growth phenomena, $e^{1/2}$ law and discussion}

For many natural phenomena with restricted growth process, the number $f(t)$
in growth first increases, reaching to its maximum growth rate at some time
(denoted $t=L$), and finally decays after the maximum occurs (at time $t=D$%
). Examples include measures of size of living tissues (length, skin area
and weigth) and spreading of epidemics in biology, restricted population
growth and income distributionn in social science, expansion of city size,
spray process in technology, [1] etc. \ 

Here, together with the analysis and data of two previous papers [2]-[3]
where however the generality and simplicity of the result were not revealed
(at least the number $e$ is not found in the formulas published there), we
want to show, though we prefer not to use prove, that the ratio between the
time for maximum and the time for maximum growth rate satisfies the
following $e^{\frac{1}{2}}$ law

\begin{equation}
\frac{D}{L}=e^{\frac{1}{2}}\text{ or }\frac{D^{2}}{L^{2}}=e\text{ or }\ln 
\frac{D}{L}=\frac{1}{2}\text{ \ (}e^{\frac{1}{2}}\text{ law)}  \label{eq0}
\end{equation}

Below are important discussions.

\textbf{Simplicity and generality}. \ The ratio expression (\ref{eq0}) is
only related to $e$, the natural exponent 
\begin{equation*}
e=2.7182\cdots
\end{equation*}%
This is elegant and simple. Does this add to the mystry [4]-[5] of this
number? On the log scale this ratio becomes one half, the simplicity of
which derserves further imagination. Most importantly, the two
characteristic time scales $L$ and $D$ should be problem dependent according
to our common sense, while according to (\ref{eq0}) their ratio is problem
independent, without any other free parameters needing ajustment. Does this
mean that simplicity is always associated with complexity [6]?

\textbf{Simple way to obtain the results}. In references [2]-[3], this
problem has been studied for spreading of droplet size and for spreading of
epidemics. Though the basic method for analysis and for obtaining the
results has been reported in [2]-[3], there is no remark about the
generality, notably, the ratio $L/D$ is not considered in [2] while it is
expressed as $L/D=1.\,\allowbreak 649$ in [3], without noticing that it can
be expressed in the elegant form of (\ref{eq0}). In the next section we
shall repeat the analysis to obtain (\ref{eq0}), as if for a more general
problem, rather than just for spray process as in [2], or epidemic spreading
as in [3]. Moreover, we shall use more experimental data to support the
intermediate theoretical parameters during the derivation for (\ref{eq0}).
The basic idea is to assume that $f(t)$, which can be expressed as the
log-normal distribution, is associated with an entropy, and for restricted
growth problem, the way for restriction is such that the rate of production
of entropy is maximized.

\textbf{Validation and agreement}. \ It is remarkable that the SARS data in
the years of 2003 obeys $L/D=1.\,\allowbreak 649$ thus (\ref{eq0}),
according to reference [3]. The reported hospitalized cases for each city
obeys this rule individually. Though this ratio is not considered for
droplet production in reference [2], the agreement between the experimental
data and the theoretical curve $f(t)$ mean that (\ref{eq0}) is also
satisfied, since (\ref{eq0}) is implicated in the theoretical curve $f(t)$.
Another rough example appear to be the height of humain body. For boys the
maximum height occurs near 23 years old while the maximum growth rate is at
the age near 14. If we use these data (the data could not be exact, and
sometimes a range is given, see [7] ), the ratio is close to $e^{-1/2}$. \
Validations against other types of growth problem are necessary to assess
further the generality of (\ref{eq0}). However, data were not reported as we
require and we have to leave further comparisons to those we have
accumulated the data for the specific problems they consider. It is
interesting to know counter examples. The spreading of droplet size and the
spreading of SARS are however quite different problems and the fact that
they obey (\ref{eq0}) would mean the possiblity of generality.

\textbf{How to use }(\ref{eq0}). It would be just for enjoy to see if for a
production process the law (\ref{eq0}) is obeyed. For instance the growth of
height of humain body happens to comply with it roughly. One may wonder how
it is possible that this law is independent of problem. However, if this law
is incorrect, the departure between it and the observation of the above
three types of examples would be extremely great! \ If this could be further
assessed by many other examples, then we may wish to use it to predict the
(restricted) growth, notably the time $D$ for maximum, after the inflextion
point (L) appears. Howver, even this is indeed so, there is a problem to
know the initial time for production (that is $L$, that should be counted
from the first time production occurs, is unknown even though we know the
day at which the maximum growth rate occurs). For instance, if this is an
epidemic, then it is very hard to know the first day that the epidemic
starts. This, however, could be evaluated as follows. At the inflexion
point, we record the number $f(L)$ and its growth rate $df(L)dt$, then $L$
can be computed as (see next section)%
\begin{equation}
L=\frac{3f(L)}{\left. \frac{df(t)}{dt}\right\vert _{t=L}}  \label{eq0b}
\end{equation}

\section{ The way to obtain the results}

As remarked above, most materials have appeared in \ references [2]-[3],
though not considered as a general way and the result is not expressed in
its elegant form (\ref{eq0}). Here we consider the problem to be universal,
repeat the essential analysis, reexpress the results in its form as stated
in section 1, and use more data to support the intermediate parameter
required to achieve finally (\ref{eq0}) and (\ref{eq0b}).

Many natural phenomena in growth can be described by the log-normal
distribution [1],

\begin{equation}
\bigskip f(t)=\frac{1}{\sqrt{2\pi }\sigma t}\exp \left( -\frac{(\ln t-\mu
)^{2}}{2\sigma ^{2}}\right)  \label{eqq1}
\end{equation}%
since the growth rate is independent of size following the Gibrat's law[8].
For spreading of epidemics, and in fact the procedure is not restricted to
epidemics, it was proved that (\ref{eqq1}) is the solution of a very simple
equation for growth rate[3]. Here ${\normalsize \mu =\ln D+\sigma ^{2}}$ is
a location parameter and the standard deviation ${\normalsize \sigma }$ is
the scale parameter on the logarithmic scale. For a particular process, $%
{\normalsize \sigma }$ is often fitted against experimental data in the
literature, without remarking whether there be some instrinsic mechanism so
that ${\normalsize \sigma }$ must take some specific value.

In fact, the standard deviation ${\normalsize \sigma }$ characterizes the
width of $f(t)$. The internal growth mechanism would make the curve $f(t)$
wider and wider. If the growth process is unrestricted, then $f(t)$ will
continue to grow without a maximum. For instance, for a highly communicable
epidemic such as SARS, if there is no restriction such as publication
intervention, then the majority of population will be infected. If there is
no restriction, then a city size will continue to grow. The restriction
effect provides a dissipation mechanism to prevent the curve to become
infinitely wide. The width cesses to increase when the maximum dissipation
rate is reached. According to reference [9], maximum dissipation rate
corresponds to maximum entropy production. This is the so-called maximum
entropy production principle, though questioned as a general principle but
often found correct and useful for determining parameters.

\bigskip Use the Shannon entropy $S(\sigma )=-\int_{0}^{\infty
}t^{1-3}f(t)\ln \left( t^{1-3}f(t)\right) dt^{3}$. It is found that [2] for $%
f(t)$ defined by (\ref{eqq1}), 
\begin{equation*}
S(\sigma )=3\left( \ln \left( \sqrt{2\pi }\sigma \right) +3\left( 
{\normalsize \ln D+\sigma ^{2}}\right) +\frac{1}{2}\right)
\end{equation*}%
Maximizing the entropy production rate, that is setting $\frac{d^{2}S(%
{\normalsize \sigma })}{d{\normalsize \sigma }^{2}}=0$, yields

\begin{equation}
\sigma =\frac{\sqrt{6}}{6}  \label{epp2}
\end{equation}%
In reference [3], it is the value $\sigma =0.401$ that is used. For spray
processs, references [10],[11] and [12] fitted the experimental data to give
the following range of $\sigma $,

\begin{equation}
\sigma =\left( 0.977,1.71\right) \frac{\sqrt{6}}{6}\text{, \ \ }\sigma
=\left( 1.009,1.065\right) \frac{\sqrt{6}}{6}\text{ and \ }\sigma =1.1023%
\frac{\sqrt{6}}{6}  \label{eqpp3}
\end{equation}%
The comparison in references [2]-[3] with experimental data and the close
agreement between (\ref{eqpp3}) and (\ref{epp2}) support the theoretical
value of (\ref{epp2}).

\bigskip Now for $f(t)$ defined by (\ref{eqq1}), we have

\begin{equation}
\frac{d^{2}f(t)}{dt^{2}}=\allowbreak g(t)\left( \sigma ^{2}\ln \frac{t}{D}%
-\sigma ^{2}+\ln ^{2}\frac{t}{D}\right) \allowbreak  \label{eqq4}
\end{equation}%
where%
\begin{equation*}
g(t)=\frac{1}{2}\frac{\sqrt{2}}{\sqrt{\pi }t^{3}\sigma ^{5}}\exp \left( -%
\frac{1}{2\sigma ^{2}}\left( \ln \frac{t}{D}-\sigma ^{2}\right) ^{2}\right)
>0
\end{equation*}

By definition of inflexion point, i.e., $t=L$ at which $\frac{d^{2}f(L)}{%
dt^{2}}=0$, we obtain from (\ref{eqq4}) the following relation 
\begin{equation*}
\sigma ^{2}\ln \frac{L}{D}-\sigma ^{2}+\ln ^{2}\frac{L}{D}=0
\end{equation*}
which can be solved to give%
\begin{equation}
\frac{D}{L}=e^{\frac{1}{2}}\text{ or }\frac{D^{2}}{L^{2}}=e  \label{eqq5}
\end{equation}%
where $e=2.71828\cdots $ is the natural exponential!

To determine the initial date for regular production, or to know the number $%
L$ (cumulated days to reach the inflexion point counting from the initial
date), we use (\ref{eqq1}) \ to write 
\begin{equation*}
\left. \frac{df(t)}{dt}\right\vert _{t=L}=-\frac{1}{\sigma ^{2}L}f(L)\ln 
\frac{L}{D}
\end{equation*}
Hence 
\begin{equation}
L=\left( \frac{1}{2}+\frac{1}{2}\sqrt{\frac{4}{\sigma ^{2}}+1}\right) \frac{%
f(L)}{\left. \frac{df(t)}{dt}\right\vert _{t=L}}=\frac{3f(L)}{\left. \frac{%
df(t)}{dt}\right\vert _{t=L}}  \label{eqll}
\end{equation}

Hence we have given all the required derivaion for what is needed in section
1.


\begin{thebibliography}{99}
\bibitem{} http://en.wikipedia.org/wiki/Log-normal\_distribution

\bibitem{} Wu Z.N. (2003) Prediction of the size distribution of secondary
ejected droplets by crown splashing of droplets impinging on a solid wall,
Probablistic Engineering Mechanics, vol 18, pp. 241-249.

\bibitem{} Wang W.B., Wang C.F., Wu Z.N. and Hu R.F., 2013 Modelling the
spreading rate of controlled communicable epidemics through an entropy-based
thermodynamic model. SCIENCE CHINA Physics,Mechanics \& Astronomy,
Doi:10.1007/s11433-013-5321-0

\bibitem{} http://en.wikipedia.org/wiki/E\_(mathematical\_constant)

\bibitem{} Eli Maor (2009) e : The Story of a Number, Princeton University
Press, Princeton

\bibitem{} She Z.S., (2012), A New Framework for Complex System, Science
Presss (China). 

\bibitem{} http://en.wikipedia.org/wiki/Human\_height

\bibitem{} Sutton, J. (1997), "Gibrat's Legacy", Journal of Economic
Literature XXXV, 40--59.

\bibitem{}  Ziegler H. (1983),An Introduction to Thermomechanics,
North-Holland Publ. Company, Amsterdam. 

\bibitem{} Stow CD, Stainer RD. The physical products of a splashing water
drop. J Met Soc Japan 1977;55:518--31.

\bibitem{} Samenfink W, Elsaper A, Dullenkopf K, Wittig S. Droplet
interaction withshear-driven liquid films: analysis of deposition and
secondary droplet characteristics. Int J Heat Fluid Flow 1999;20:462--9.

\bibitem{} Schmehl R, Rosskamp H, Willmann M, Wittig S. CFD analysis of
spray propagation and evaporation including wall film formation and
spray/film interactions. Int J Heat Fluid Flow 1999;20:520--9.
\end{thebibliography}
\end{document}